\documentclass[preprint]{elsarticle}

\usepackage{amsmath}
\usepackage{amssymb}
\usepackage{xspace}
\usepackage{mathbbol}
\usepackage{tikz}
\usepackage{algorithm2e}
 
\usepackage{enumerate}

\newtheorem{theorem}{Theorem}[section]
\newtheorem{lemma}[theorem]{Lemma}
\newtheorem{proposition}[theorem]{Proposition}

\newtheorem{definition}[theorem]{Definition}
\newtheorem{remark}[theorem]{Remark}

\newcommand \Nset {\ensuremath{\mathbb{N}}\xspace}

\def\endproof{\hfill$\Box$ \par}

\begin{document}


\begin{frontmatter}

\title{Generalized Huffman Coding for Binary Trees with Choosable Edge Lengths}
\author{Jens Ma{\ss}berg}
\address{Institute of Optimization and Operations Research, University of
Ulm, Germany, jens.massberg@uni-ulm.de}


\begin{abstract}
In this paper we study binary trees with choosable edge lengths, in
particular rooted binary trees with the property that the two edges leading
from every non-leaf to its two children are assigned integral lengths
$l_1$ and $l_2$ with $l_1+l_2 =k$ for a constant $k\in\mathbb{N}$.
The depth of a leaf is the total length of the edges of the unique
root-leaf-path.
 
We present a generalization of Huffman Coding that can decide 
in polynomial time if for given values $d_1,\ldots,d_n\in\mathbb{N}_{\geq
0}$ there exists a rooted binary tree with choosable edge lengths with $n$
leaves having depths at most $d_1,\ldots ,d_n$.
\end{abstract}


\end{frontmatter}

\section{Introduction}

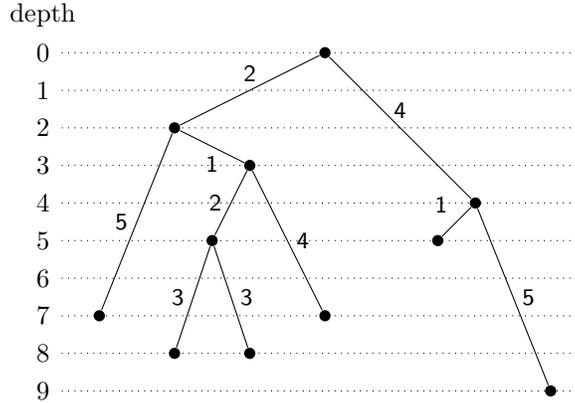
\begin{figure}[ht]
\begin{center}
 \begin{tikzpicture}[scale=0.5]
   \foreach \y in {0,...,9} 
   {\draw[dotted] (-7,10-\y) -- (7,10-\y);
   \node at (-7.5,10-\y) {\y};
   } 
   \node at (-7.5,11) {depth};
   
   \node[circle,fill, inner sep=1.5] at (0,10) (a) {};
   \node[circle,fill, inner sep=1.5] at (-4,8) (a1) {}; 
   \node[circle,fill, inner sep=1.5] at (4,6) (a2) {};

   \node[circle,fill, inner sep=1.5] at (3,5) (a21) {};
   \node[circle,fill, inner sep=1.5] at (6,1) (a22) {};

   \node[circle,fill, inner sep=1.5] at (-6,3) (a12) {};
   \node[circle,fill, inner sep=1.5] at (-2,7) (a11) {};
   \node[circle,fill, inner sep=1.5] at (0,3) (a111) {};
    \node[circle,fill, inner sep=1.5] at (-3,5) (a112) {};

      \node[circle,fill, inner sep=1.5] at (-4,2) (a1121) {};
    \node[circle,fill, inner sep=1.5] at (-2,2) (a1122) {};
  

    \path[every node/.style={font=\sffamily\small}]
    (a) edge node[above] {2} (a1)
    edge node [above] {4} (a2)
    (a1) edge node[left] {5} (a12)
    edge node[below] {1} (a11)
    (a11) edge node [right] {4} (a111)
    edge node [left] {2} (a112)
    (a112) edge node[left] {3} (a1121)
    edge node[right] {3} (a1122)
    (a2) edge node [above left] {1} (a21)
    edge node [right] {5} (a22);

 \end{tikzpicture}
  \caption{An ${\cal L}'(6)$-tree for the leaf signature $\{5,7,7,8,8,9\}$.}
 \label{fig1}
\end{center}
\end{figure}

For a fixed $k\in\Nset$ we define
${\cal L}'(k)=\{ \{ i,k-i\} \mid 1\leq i \leq k-1, i\in \mathbb{N}\}.$
An ${\cal L}'(k)$-tree is a rooted strict binary tree with the
property that the two edges leading from every non-leaf to its two children are
assigned lengths $l_1$ and $l_2$ with $\{l_1,l_2\}\in {\cal L}'(k)$. In this
context we will from now on call a multi-set  $D=\{d_1,\ldots, d_n\}$ a
\emph{leaf signature}.
In this paper we show for fixed $k$ how to decide in polynomial time for a given
leaf signature $\{d_1,\ldots,d_n\}$ if there
exists an ${\cal L}'(k)$-tree with $n$ leaves at depths at most $d_1,\ldots,
d_n$ (in any order) and how to construct such a tree.
See Figure \ref{fig1} for an example of an ${\cal L}'(6)$-tree for the leaf
signature $\{5,7,7,8,8,9\}$. 
As the numbers of a leaf signature give an upper bound on the depths of the leaves,
the tree of the  example in Figure \ref{fig1} is also an ${\cal L}'(6)$-tree for
the leaf 
signatures $\{7,7,7,9,9,9\}$ and $\{9,9,9,9,9,9\}$.

If $k=2$, then all edges of an ${\cal L}'(k)$-tree have length $1$. In this
case we get classical binary trees where the depth of a leaf is equal to the
number of
edges of the unique path from the root to the leaf. By Kraft's inequality
\cite{kraft} an ${\cal L}'(2)$-tree for $D$ exists if and only if 

\begin{equation}
 \sum_{i\in\{1,\ldots, n\}} 2^{-d_i} \leq 1.
\end{equation}

Such a tree can be constructed using the Huffman Coding algorithm
\cite{huffman1952method}.

${\cal L}'(k)$-trees have first been considered by Ma{\ss}berg and
Rautenbach \cite{MassbergRautenbach}. They are motivated by
the so-called repeater tree problem,
a problem that occurs in VLSI-design.
Repeater trees are tree structures consisting of wires and internal gates.
They are required to distribute an electronical signal from a root to
several locations on a chip, called sinks, while not exceeding individual time
restrictions.
In general there is a certain degree of freedom to distribute delay to
different branches of the tree.
By inserting a gate at a vertex of the tree it is possible to reduce the
delay of one of the branches while increasing the delay on the other branch
by about the same amount.
As there are only a discrete number of gates with different sizes available,
this effect can be modeled by the ${\cal L}'(k)$-trees:
The lengths of the edges correspond to the delays on these edges and the $d_i$
correspond to the required arrival times of a sink $i$.
Then the task is to build an ${\cal L}'(k)$-tree such that the signal reaches
each sink on time. 
For more details on the repeater tree problem and its connection to binary
trees with choosable edge lengths we refer the reader to
\cite{a5} and \cite{heldrotter}.

In \cite{MassbergRautenbach} it has been shown how to construct ${\cal
L}'(4)$-trees in polynomial time.
For $k>4$ it was an open problem if there exists a polynomial algorithm that
can decide the existence of ${\cal L}'(k)$-trees for given instances.
We show that we can decide the existence in time $O(n^{k+3})$, that is, in
polynomial time for constant
$k$.

Our problem is also related to prefix-free codes with unequal letter costs
(see e.g. \cite{br,gilbert,golin,go2}).
Nevertheless, there are significant differences between these problems.
First, in our problem we ask for a tree satisfying given depth restrictions
for the leaves while for prefix-free codes the task is to find a tree minimizing
$\sum_{i\in\{1,\ldots,n\}}\text{depth}(i)w_i$ where $\{w_1,\ldots,w_n\}$ are
given numbers and depth$(i)$ denotes the depth of leaf $i$, $i\in\{1,\ldots,
n\}$.
Moreover, in our problem we have the freedom to choose the edge lengths
from a discrete set of numbers as long as the sum of the lengths of the edges
leaving a vertex equals $k$.


\section{Core Algorithm}

Let $D=\{d_1,\ldots, d_n\}$ be a leaf signature.
We want to decide if there exists an  ${\cal L}'(k)$-tree with $n$ leaves at
depths at most $d_1,\ldots, d_n$. If such a tree exists we call the leaf
signature $D$
\emph{realizable}.



In this section we present the core algorithm that can decide the
existence of an ${\cal L}'(k)$-tree for a given leaf signature.
The idea of the algorithm is the following:
We can reduce a leaf signature of length $n$ by combining two leaves of a
potential tree into one slightly higher up in the tree. This leads to a set of
leaf signatures of length $n-1$. If one of them is realizable, the original one
is realizable, too. By iterative application of this method we get a set of leaf
signatures of length $1$ where the existence of an ${\cal L}'(k)$-tree can be
checked easily.
In Section \ref{sec3} we refine the algorithm to get a polynomial running time.

For two integers $a_1,a_2\in \mathbb{Z}$  we define

\begin{equation}
 \omega_k(a_1,a_2) := \min\{a_1,a_2\} - \max\left\{1, \left\lceil
\frac{k-|a_1-a_2| }{2}
 \right\rceil\right\}.
\end{equation}

The idea of the core algorithm relies on the following observation.
\begin{proposition}\label{prop1}
A leaf signature $D=\{d_1,\ldots, d_n\}$, $n\geq 2$, is realizable if and only
if there exist $i,j$ with $1\leq i < j \leq n$ such that
$D\setminus \{d_i,d_j\} \cup \{\omega_k(d_i,d_j)\}$
is realizable.
\end{proposition}
{\it Proof:}
 Assume that $D$ is realizable and let $T$ be an ${\cal L}' (k)$-tree realizing
 $D$.
 Then there must be two leaves $v$ and $w$ of $T$ that have a common parent $u$.
 Let $d_i$ and $d_j$, $i<j$, be the depth limits from $D$ assigned to $v$ and
$w$.

Let $d:V(T)\rightarrow \mathbb{N}$ be the depth of the vertices of $T$ with
respect to the length of the edges.
We show that $d(u)\leq\omega(d_i,d_j)$.
 As $T$ is an ${\cal L}' (k)$-tree realizing $D$ we have $d(u) \leq d(v)-1\leq
d_i-1$, $d(u) \leq d(w)-1\leq d_j-1$ and
 $k=(d(v) -d(u)) + (d(w) - d(u)) \leq d_i+d_j-2d(u)$.
 We conclude, using that $d(u)\in\mathbb{N}$:

 \begin{equation*}
   d(u) \leq \min\{d_i,d_j\}-1
 \end{equation*}

 and 

 \begin{equation*}
   d(u) \leq \left\lfloor\frac{d(v)+d(w)-k}{2}\right\rfloor = \min\{d_i,d_j\} -
\left\lceil \frac{k-|d_i-d_j|}{2} \right\rceil.
 \end{equation*}
Therefore we have $d(u)\leq \omega_k(d_i,d_j)$ and the tree
$(V(T)\setminus\{v,w\},E(T)\setminus\{uv,uw\})$ realizes $D\setminus \{d_i,d_j\}
\cup \{\omega_k(d_i,d_j)\}$.

 On the other hand assume, that there are $i,j\in\{1,\ldots,n\}$, $i<j$, such that 
 $D'=(D\setminus\{d_i,d_j\})\cup\{\omega_k(d_i,d_j)\}$ is realizable.
 Let $T$ be an ${\cal L}' (k)$-tree realizing $D'$ and let $v$ be the leaf of
 $T$ assigned to $\omega_k(d_i,d_j)$.
 Add two edges of length $a=\max\{k-1, d_i-\omega_k(d_i,d_j)\}$ and $b=k-a$ at $v$.
 Then the two newly inserted leaves have depth at most $d_i$ and $d_j$ and thus
 the new tree realizes $D$.
\endproof

By repeatedly applying Prop. \ref{prop1} we get to the algorithm outlined at the
beginning of the section. Unfortunately, the number of leaf signatures that are
computed grows exponentially in the length of the initial leaf signature.
\section{Refined Algorithm}\label{sec3}

In this section we refine the algorithm outlined in the previous section
in order to get a polynomial running time. The refinement depends on the 
observations that the
values of a leaf signature cannot be too big and that we can restrict ourselves to
leaf signatures identical to the original signature except for the
bottom $k$ layers.

First we introduce the notion of domination which will be useful in the proofs.
\begin{definition}
 Let $A=\{a_1,\ldots, a_n\}$ and $B=\{b_1,\ldots,b_n\}$ be two leaf signatures
of
the same length. If there exists a permutation $\pi:\{1,\ldots,n\}\rightarrow
\{1,\ldots,n\}$ such that $a_i\leq b_{\pi(i)}$ for all $i\in\{1,\ldots,n\}$
then $A$ is \emph{dominated} by $B$.
\end{definition}

Obviously, it is sufficient to consider only sets of leaf signatures, where no 
signature is dominated by another.


Next we note that we can assume the values of a leaf signature not to be too
big.
\begin{proposition}\label{max_el}
The leaf signature $D=\{d_1,\ldots, d_n\}$ is realizable if and only if the leaf
signature $D'=\{\min\{d_1,(k-1)(n-1)\},\ldots, \min\{d_n,(k-1)(n-1)\}\}$ is
realizable.
\end{proposition}
{\it Proof:}
Assume there is an ${\cal L}' (k)$-tree $T$
realizing $D$.
As $T$ contains $n$ leaves,  every root-leaf-path consists of at most $n-1$ edges.
Moreover, in an ${\cal L}' (k)$-tree every edge has length at most $k-1$.
We conclude that the leaves of $T$ have depth at most
$\min\{d_1,(k-1)(n-1)\},\ldots, \min\{d_n,(k-1)(n-1)\}$ and therefore $D'$ is
realizable.

The other direction of the proof follows from the fact, that $D'$ is dominated
by $D$.
\endproof

Now we show that we can assume each computed leaf signature to be identical to
the original signature except for the bottom $k$ layers.

\begin{proposition}\label{prop:3}
 The leaf signature $D=\{d_1,\ldots, d_n\}$ is realizable if and only if there
exist
$i,j$ with $1\leq i < j \leq n$ such that the signature
  \begin{equation}\label{eq:omega}
  \big(\{\min(d_1,\omega),\ldots, \min(d_n,\omega)
\} \setminus
\{\min(d_i,\omega),\min(d_j,\omega)\}\big) \cup\{\omega_k(d_i,d_j)\}
 \end{equation}
with $\omega = \omega_k(d_i,d_j)+k-1$ is realizable.

\end{proposition}
{\it Proof:}
 Let $T$ be an ${\cal L}' (k)$-tree realizing $D$,
 let $u$ be an internal vertex of $T$ of maximum depth and denote by $v,w$ the
 two children of $u$.
 Obviously, $v$ and $w$ are leaves of $T$. Let $d_i$ and $d_j$, $1\leq i < j
 \leq n$, be the values assigned to $v$ and $w$.
 By the proof of Prop. \ref{prop1} we have $d(u)\leq \omega_k(d_i,d_j)$.
 As $u$ is an internal vertex of maximum depth and all edges in $T$ have length
 at most $k-1$, all leaves of $T$ have depth of at most $d(u)+k-1\leq
 \omega_k(d_i,d_j)+k-1$.
 Deleting the leaves $v,w$ and their incident edges from $T$ we obtain a tree
realizing the signature (\ref{eq:omega}).

 On the other hand assume that there are $i$ and $j$ with $1\leq i < j \leq n$
 such that $A=(\{\min(d_1,\omega),\ldots, \min(d_n,\omega)
\}\setminus\{\min(d_i,\omega),\min(d_j,\omega)\})\cup\{\omega_k(d_i,d_j)\})$ is
realizable for $\omega=\omega_k(d_i,d_j)+k-1$. Note, that $A$ is dominated by
$A'=(\{d_1,\ldots,d_n\}\setminus\{d_i,d_j\})\cup \{\omega_k(d_i,d_j)\}$ and thus
$A'$ is
realizable. Then by Proposition \ref{prop1} the leaf signature $D$ is
realizable.
\endproof
%
%
%
%
%
%
%
%
%

The above results lead us to Algorithm \ref{algorithm2} where one after another
set $\mathcal{M}_z$ of leaf signatures of length $z$ are computed, starting
with $\mathcal{M}_n$ containing only the initial leaf signature of length $n$.

\begin{algorithm}[ht]
\LinesNumbered\SetAlgoLined
\KwIn{An instance $D=\{d_1,\ldots, d_n\}$ and $k\in\mathbb{N}, k\geq 2$.}
\KwOut{Returns true iff there exists an ${\cal L}'(k)$-tree for $D$.}
\BlankLine
$\mathcal{M}_n \leftarrow \{D\}$\;
\For{$z=n-1$ \textbf{\emph {to}} $1$ 
}
{
  $\mathcal{M}_z \leftarrow \emptyset$\;
 \ForEach{leaf signature $A=\{a_1,\ldots, 
a_{z+1}\}\in\mathcal{M}_{z+1}$\label{alg:A}
 }
 {
 
$\mathcal{B}_A\leftarrow \emptyset$\;
\ForEach{$i,j\in \{1,\ldots, z+1\}, i<j$\label{alg:loop}} {
Compute leaf signature $B$ out of $A$ by replacing $a_i$ and $a_j$
by $\omega_k(a_i,a_j)$ and truncating large values in $B$ according to
Prop. \ref{max_el} and Prop. \ref{prop:3}\;\label{alg:cut2}
    $\mathcal{B}_A \leftarrow \mathcal{B}_A\cup \{B\}$\;
  }
Remove all signatures from $\mathcal{B}_A$ that are dominated by other
signatures in $\mathcal{B}_A$\;
$\mathcal{M}_z \leftarrow \mathcal{M}_z\cup \mathcal{B}_A$\;
 }

(optional: Remove leaf signatures $A\in\mathcal{M}_z$ that are dominated by other
leaf signatures $B\in\mathcal{M}_z$\label{alg:dom})\;
}
\lIf{$\{i\}\in \mathcal{M}_1$ for some $i\geq 0$}
{\Return{{true}}\;}
{\Return{{false}}\;}
\caption{${\cal L}'(k)$-tree decision algorithm.}
\label{algorithm2}
\end{algorithm}

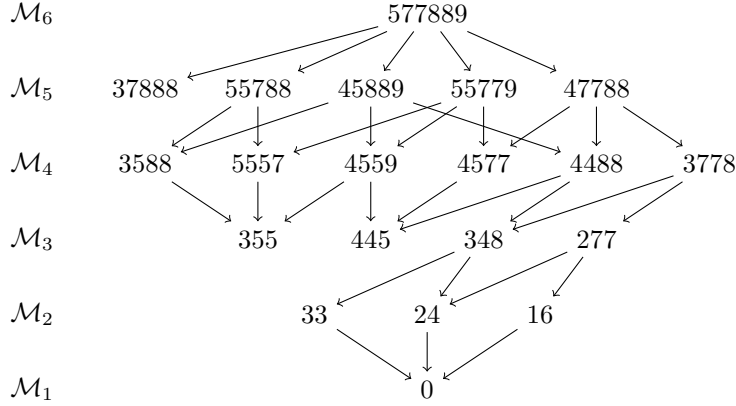
\begin{figure}[ht]
 \begin{center}
 \begin{tikzpicture}[xscale=1.5]
  \node at (0,5) (n51) {577889};

  \node at (-2-0.5,4) (n41) {37888};
  \node at (0-0.5,4) (n42) {45889};
  \node at (2-0.5,4) (n43) {47788};
  \node at (1-0.5,4) (n44) {55779};
  \node at (-1-0.5,4) (n45) {55788};
  
  \node at (-2.5,3) (n31) {3588};
  \node at (2.5,3) (n32) {3778};
  \node at (1.5,3) (n33) {4488};
  \node at (-0.5,3) (n34) {4559};
  \node at (0.5,3) (n35) {4577};
  \node at (-1.5,3) (n36) {5557};

  \node at (1.5,2) (n21) {277};
  \node at (0.5,2) (n22) {348};
  \node at (-1.5,2) (n23) {355};
  \node at (-0.5,2) (n24) {445};
  
  \node at (1,1) (n11) {16};
  \node at (0,1) (n12) {24};
  \node at (-1,1) (n13) {33};

   \node at (0,0) (n00) {0};

\draw[->] (n51) -- (n41);
\draw[->] (n51) -- (n42);
\draw[->] (n51) -- (n43);
\draw[->] (n51) -- (n44);
\draw[->] (n51) -- (n45);

\draw[->] (n43) -- (n32);
\draw[->] (n42) -- (n31);
\draw[->] (n42) -- (n34);
\draw[->] (n43) -- (n33);
\draw[->] (n43) -- (n35);
\draw[->] (n44) -- (n34);
\draw[->] (n44) -- (n36);
\draw[->] (n45) -- (n31);
\draw[->] (n44) -- (n35);

\draw[->] (n42) -- (n33);
\draw[->] (n45) -- (n36);

\draw[->] (n31) -- (n23);
\draw[->] (n32) -- (n21);
\draw[->] (n33) -- (n22);
\draw[->] (n34) -- (n23);
\draw[->] (n34) -- (n24);
\draw[->] (n35) -- (n24);
\draw[->] (n36) -- (n23);
\draw[->] (n33) -- (n24);

\draw[->] (n32) -- (n22);

\draw[->] (n21) -- (n11);
\draw[->] (n21) -- (n12);
\draw[->] (n22) -- (n13);
\draw[->] (n22) -- (n12);

\draw[->] (n11) -- (n00);
\draw[->] (n12) -- (n00);
\draw[->] (n13) -- (n00);

\node at (-3.5,5) {$\mathcal{M}_6$};
\node at (-3.5,4) {$\mathcal{M}_5$};
\node at (-3.5,3) {$\mathcal{M}_4$};
\node at (-3.5,2) {$\mathcal{M}_3$};
\node at (-3.5,1) {$\mathcal{M}_2$};
\node at (-3.5,0) {$\mathcal{M}_1$};
 \end{tikzpicture}
 \end{center}
 \caption{Computed leaf signatures for the instance
$\{5,7,7,8,8,9\}$.}
 \label{fig:example}
\end{figure}

Figure \ref{fig:example} shows the sets $\mathcal{M}_z$, $z\in\{1,\ldots, 6\}$,
computed by the algorithm for the instance $\{5,7,7,8,8,9\}$. Dominated 
signatures are removed. The arrows show where each new signature comes from.
As the final set $\mathcal{M}_0$ contains the leaf signature $\{0\}$, the
initial signature is 
realizable.

Let $D=\{d_1,\ldots, d_n\}$ be the input of the algorithm.
In order to be able to reproduce the reduction steps  we introduce a function
$p(\cdot)$ that assigns a leaf signature $B$ to the signature it replaces, that
is, $p(B)=A$ for $A$ and $B$ as in Line \ref{alg:cut2} of the algorithm.
For a leaf signature $B$ we denote by $l(B)$ the minimum value of an
element that  has been added to $B$, that is, $l(D)=\infty$ and $l(B)=
\min\{l(A),\omega(a_i,a_j)\}$ for $A,B,i,j$ as in Line \ref{alg:cut2}.
This implies that we have only removed and changed 
elements $b$ with $b>l(B)$.
Thus if $B=\{b_1,\ldots, b_z\}$ with $b_1\leq\ldots\leq b_z$ and $
d_1\leq\ldots\leq d_z$ then

\begin{equation}\label{eq:lB}
 b_i=d_i \text{ for all }i\leq \max\{j|\, b_j<l(B)\} = \max\{j|\,
d_j<l(B)\}.
 \end{equation}

The algorithm computes all necessary leaf signatures according to Prop. 
\ref{max_el} and Prop. \ref{prop:3}.
Consequently, a leaf
signature in $\mathcal{M}_{z}$ is realizable if and only if a leaf signature
in $\mathcal{M}_{z+1}$ is realizable for all $z\in\{1,\ldots, n-1\}$ proving the
correctness of the algorithm.

For simplicity of notation we set $m(A)=\max_{a\in A} a$.
We will prove now that the running time of Algorithm \ref{algorithm2} is
polynomially bounded in $n$ for fixed $k$. To this end we show that the sizes of
the sets $\mathcal{M}_z$ are polynomially bounded in $n$.
First we show that for every computed leaf signature $B$ the
largest element in $B$ is at most $k-1$ larger than the smallest element that
has been inserted into $B$ by the algorithm.
\begin{proposition} \label{marked}
If $z\in\{1,\ldots, n-1\}$, then all leaf signatures $B\in\mathcal{M}_z$ satisfy

 \begin{equation}\label{eq:mark}
   m(B) \leq l(B) +k-1.
 \end{equation}

\end{proposition}
{\it Proof:}
 For contradiction we assume that $B$ is a set of maximum cardinality 
 computed by the algorithm that contradicts (\ref{eq:mark}).
 
 Set $A=p(B)$. Obviously, $l(B)\leq l(A)$ and $m(B)\leq m(A)$. 
 If $l(B)=l(A)$ then 
 $m(A)\geq m(B) > l(B)+k-1 = l(A)+k-1$,
 i.e. $A$ also does not satisfy (\ref{eq:mark}), contradicting the maximality
 of $B$.
 
 Thus $l(B)<l(A)$. But in this case $l(B)=\omega(a_i,a_j)$ for the two elements 
 $a_i,a_j\in A$ that are replaced. 
 As all values of $B$ are truncated in Line 
 \ref{alg:cut2} of the algorithm according to Prop. \ref{prop:3},
 we obtain $m(B)\leq m \leq \omega(a_i,a_j)+k-1 =l(B)+k-1$,
 which is a contradiction and completes the proof.
\endproof

Using the previous result we show that the size of $\mathcal{M}_z$ is 
polynomially bounded in $z$.

\begin{lemma}\label{lem_mz}
 If $z\in\{1,\ldots, n-1\}$, then
 \begin{equation}
   |\mathcal{M}_z| \leq z^{k}. 
 \end{equation}
\end{lemma}
{\it Proof:}
  Define $\mathcal{M}_z^i :=\{A\in \mathcal{M}_z:\, l(A)=i\}$ for $i\in
  \mathbb{N}$.
  By Prop. \ref{max_el} and the truncation 
  of large values in Line \ref{alg:cut2}
  of the algorithm according to Prop. \ref{max_el},
  we know $\mathcal{M}_z^i = \emptyset$ for
  $i>(z-1)(k-1)$.
  
  Let $i\in\{0,\ldots, (n-1)(k-1)\}$ and $A=\{a_1,\ldots, a_z\}\in
  \mathcal{M}_z^i$ such that $a_1\leq\ldots \leq a_z$.
  Recall that $D=\{d_1,\ldots, d_n\}$, $d_1\leq\ldots\leq d_n$, is the input of
  the algorithm. 
  Set $t=\max \{j:\, d_j < l(A)\}$.  
  By the definition of $l(A)$ and (\ref{eq:lB}), we conclude

  \begin{equation}
    a_j = d_j
  \end{equation}

  for $j \in \{1,\ldots, t\}$.
  On the other hand, by Prop. \ref{marked},

  \begin{equation}
    a_j \in \{l(A),\ldots, l(A)+k-1\}  
  \end{equation}

  for $j \in \{t+1,\ldots, z\}$.

 This implies that the sets $A\in\mathcal{M}_z^i$ only differ in the largest
 $(z-t)$ elements and each of these elements can only take one of $k$ different
 values.
 Thus the number of different sets $A$ (without removing dominated ones)
 is at most the number of 
 integral partitions of $z-t$ into the
 sum of $k$ non-negative integers.
 This number equals $\tbinom{z-t+k-1}{k-1}$ and is bounded by
$\frac{(z-t)^{k-1}}{k-1}\leq \frac{z^{k-1}}{k-1}$.

 Altogether we have at most $1+(z-1)(k-1)$ sets $\mathcal{M}_z^i$
 that are non-empty and each of these sets has at most $z^{k-1}$
 elements. Hence

 \begin{equation}
  |\mathcal{M}_z| = \sum_{i\in \mathbb{N}}|\mathcal{M}^i_z| 
  \leq (1+ (z-1)(k-1))\frac{z^{k-1}}{k-1} 
  \leq z^{k}.
 \end{equation}
 
 This finishes the proof.
\endproof

Before we can prove the running time of the algorithm we show that 
for each leaf signature $A$ the size of the set $\mathcal{B}_A$ is not too big.

\begin{lemma}\label{lemma:MA}
 For any leaf signature $A=\{a_1,\ldots, a_{z+1}\}$ the set 
 $\mathcal{B}_A$ contains at most $k\cdot z$ elements.
Moreover, the set $\mathcal{B}_A$ can be computed in time $O(kn^2)$.
\end{lemma}
{\it Proof:}
 W.l.o.g. assume $a_1\leq\ldots \leq a_{z+1}$.
 For every pair $i,j\in\{1,\ldots, z+1\}$, $i\leq j$, a new signature $B$ is
 computed out of $A$ in the \textbf{foreach} loop in Line \ref{alg:loop}. We
 denote this signature by $B_{i,j}$.
 
 First note that 
$\omega_k(a_i,a_j)=a_i-\max\{1, \lceil
(k-a_j+a_i)/2\rceil \}$ and thus 
\begin{equation}
a_i-k+1\leq a_i-\lceil k/2\rceil\leq
\omega_k(a_i,a_j) \leq a_i-1.
\end{equation}

Now assume that there exist indices $i,j,j'$, $1\leq i\leq j \leq j'\leq z+1$,
such that $\omega=\omega_k(a_i,a_j)=\omega_k(a_i,a_{j'})$. Then
$B_{i,j}=(B_{i,j'}\setminus\{\min\{a_{j'},(z-1)(k-1),\omega+k-1\}\})
\cup \{\min\{a_{j},(z-1)(k-1),\omega+k-1\} \}$, that is,
 $B_{i,j}$ is dominated by $B_{i,j'}$.
We conclude that the maximum number of signatures in $\mathcal{B}_A$ depends on
the number of possible values for $a_i$ and $\omega$, respectively.

Instead of traversing all values for $i$ and $j$ in order to compute
$\mathcal{B}_A$ it suffices to traverse all possible values of $i$ and
$\omega$. In order to construct a new signature for given $i$ and $\omega$ we
have to traverse $z+1$ elements, implying the total running time of
$O(kz^2)$.
\endproof

Joining the previous results together we are able to prove that the running 
time is polynomially bounded in the size of the input.

\begin{theorem}
For a given leaf signature $D=\{d_1,\ldots, d_n\}$ it can be decided in
time $O(n^{k+3})$ if there exists an ${\cal L}'(k)$-tree with $n$ leaves at
depth at most $d_1,\ldots, d_n$. Moreover, such a tree can be constructed in 
the same running time.
\end{theorem}
{\it Proof:}
To achieve this running time we combine Algorithm \ref{algorithm2} and the
idea of Lemma \ref{lemma:MA} in order to compute the sets $\mathcal{B}_A$.
By Lemma \ref{lemma:MA} each set $\mathcal{B}_A$ can be computed in time
$O(kz^2)$ for $z+1=|A|$. Applying Lemma \ref{lem_mz} the total running time is bounded by

 \begin{equation}
  O\Big(\sum_{z\in\{1,\ldots, n-1\}} kz^2|\mathcal{M}_z| \Big)
\subseteq O\Big(
 \sum_{z\in\{1,\ldots, n-1\}} kz^{k+2} \Big)
\subseteq O\left( n^{k+3}\right).
 \end{equation}

 As we have seen before, there exists an  ${\cal L}'(k)$-tree realizing the
 signature $D$  if and only if  $\{i\}\in \mathcal{M}_1$ for an $i\geq
0$. This tree can be constructed using the predecessor function $p(\cdot)$.
\endproof

In practice the running time can be improved by removing dominated signatures.

\begin{remark}
By removing dominated signatures after each iteration of the
main loop (see Line \ref{alg:dom} of Alg. \ref{algorithm2}),
the cardinalities of the sets $\mathcal{M}_z$ and the
running time of the algorithm can be reduced significantly in practice.
Nevertheless, it seems that the theoretical worst case running times do not
decrease in general.
\end{remark}

\section{Conclusion and Future Work}

In this paper we have presented an algorithm building rooted binary trees with choosable 
edge lengths in polynomial time for fixed $k$.
This algorithm can be seen as a generalization of Huffman coding:
If $k=2$ we are in the case of ordinary binary trees with all edges having
length $1$. 
In this case it is easy to show that for any leaf signature
$A=\{a_1,\ldots,a_z\}$ the set $\mathcal{B}_A$ only contains the signature we
get by replacing the largest two elements $a_i,a_j$, $1\leq i < j < z$, by
$\omega_2(a_i,a_j)=\min\{a_i,a_j\}-1$ (after
removing dominated signatures).
Thus $|\mathcal{M}_z|=1$ for all $z\in\{1,\ldots, n\}$ and the algorithm is
equivalent to Huffman coding.

It is still open if there is an algorithm for the 
${\cal L}'(k)$-tree with a significantly better running time, for example 
an algorithm with a running time that is polynomially bounded not only in $n$ 
but also in $k$.

\bibliography{massberg_bib}{}

\begin{thebibliography}{1}

\bibitem{a5}
C.~Bartoschek, S.~Held, J.~Ma{\ss}berg, D.~Rautenbach, and J.~Vygen.
\newblock The repeater tree construction problem.
\newblock {\em Information Processing Letters}, 110:1079--1083, 2010.

\bibitem{br}
P.~Bradford, M.~Golin, L.~L. Larmore, and W.~Rytter.
\newblock Optimal prefix-free codes for unequal letter costs and dynamic
  programming with the monge property.
\newblock {\em Journal of Algorithms}, 42(2):277--303, 2002.

\bibitem{gilbert}
E.~N. Gilbert.
\newblock Coding with digits of unequal cost.
\newblock {\em IEEE Transactions on Information Theory}, 41(2):596--600, 1995.

\bibitem{golin}
M.~Golin, C.~Mathieu, and N.~E. Young.
\newblock Huffman coding with letter costs: A linear-time approximation scheme.
\newblock {\em SIAM Journal on Computation}, 41(3):684--713, 2012.

\bibitem{go2}
M.~Golin and G.~Rote.
\newblock A dynamic programming algorithm for constructing optimal prefix-free
  codes for unequal letter costs.
\newblock {\em IEEE Transactions on Information Theory}, 44(5):1770--1781,
  1998.

\bibitem{heldrotter}
S.~Held and S.~Rotter.
\newblock Shallow-light {S}teiner arborescences with vertex delays.
\newblock {\em Integer Programming and Combinatorial Optimization}, pages
  229--241, 2013.

\bibitem{huffman1952method}
D.A. Huffman.
\newblock A method for the construction of minimum-redundancy codes.
\newblock In {\em Proceedings of the IRE}, volume~40, pages 1098--1101, 1952.

\bibitem{kraft}
L.G. Kraft.
\newblock A device for quantizing, grouping, and coding amplitude modulated
  pulses.
\newblock Master's thesis, MIT, Cambridge, 1949.

\bibitem{MassbergRautenbach}
J.~Ma{\ss}berg and D.~Rautenbach.
\newblock Binary trees with choosable edge lengths.
\newblock {\em Information Processing Letters}, 109(18):1087--1092, 2009.

\end{thebibliography}
\bibliographystyle{plain}

\end{document}